\documentclass[a4paper]{jacow}
\usepackage{pdfpages,multirow,multicol,ragged2e}  
\usepackage{graphicx}
\usepackage{subfigure}
\usepackage{float}
\usepackage{tabularx}
\usepackage{hyperref}
\usepackage[compact]{titlesec}
\usepackage{graphicx} 
\usepackage{subcaption} 
\usepackage{caption} 
\usepackage{graphicx} 
\usepackage{array} 
\usepackage{geometry} 
\begin{document}
\title{Optics tuning simulations for \NoCaseChange{{FCC-ee}} using Python Accelerator Toolbox 
\thanks{\NoCaseChange{Appears in the proceedings of the 14th International Computational Accelerator Physics Conference (ICAP’24), 2-5 October 2024, Germany.}}}

\author{E. Musa\thanks{elaf.musa@desy.de}, I. Agapov, Deutsches Elektronen Synchrotron, Hamburg, Germany \\
  T.K. Charles, ANSTO - Australian Synchrotron, Melbourne, Australia
		}

\maketitle

\begin{abstract}
\justify 

The development of ultra-low emittance storage rings, such as the e+/e- Future Circular Collider (FCC-ee) with a circumference of about 90 km, aims to achieve unprecedented luminosity and beam size. One significant challenge is correcting the optics, which becomes increasingly difficult as we target lower emittances. In this paper, we investigate optics correction methods to address these challenges. We examined the impact of arc region magnet alignment errors in the baseline optics for the FCC-ee lattice at Z energy. To establish realistic alignment tolerances, we developed a sequence of correction steps using the Python Accelerator Toolbox (PyAT) to correct the lattice optics, achieve the nominal emittance, Dynamic Aperture (DA), and in the end, the design luminosity. The correction scheme has been recently optimized and better machine performance demonstrated. A comparison was conducted between two optics correction approaches: Linear Optics from Closed Orbits (LOCO) with phase advance + $\eta_x$ and coupling Resonance Driving Terms (RDTs) + $\eta_y$. The latter method demonstrated better performance in achieving the target emittance and enhancing the DA.

\end{abstract}

\section{Introduction}

\begin{figure}[h]
       \centering
       \includegraphics*[width=0.5\textwidth]{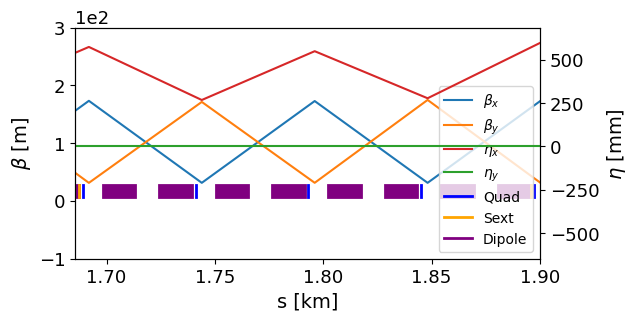}
       \caption{Section of the arc region indicating the lattice and optics parameters for the
baseline lattice. Quadrupoles, sextupoles and dipoles (rectangulars) are
shown in blue, orange and purple respectively.}
\label{fig:arc_optics1}
\end{figure}

The proposed FCC-ee \cite{fcceecdr}, 
aims to push the limits of the achievable  luminosity.
This will make the FCC-ee a unique precision instrument to study the heaviest known
particles (Z, W, H bosons and the top quark), offering great insights into new physics 
\cite{fcceephy, Blondel2019FCCeeYQ}. The FCC-ee lattices are designed to have eight arcs. The optics of the arc region for the baseline lattice rely on a FODO cell structure as illustrated in Fig.~\ref{fig:arc_optics1}. Table~\ref{tab:parameterstable} summarises some of the main parameters for the baseline optics of FCC-ee at $t \bar{t}$ and $Z$ energies.

\begin{table}[h]
\caption{Parameters of the FCC-ee baseline lattices at $t \bar{t}$ and $Z$ energies \cite{parametersTable23}.}

\centering
\small
\setlength{\tabcolsep}{4pt} 
\renewcommand{\arraystretch}{1.2} 
\begin{tabular}{|c|c|c|}
\hline
\textbf{Lattice parameter} & \textbf{Z} & \textbf{$t \bar{t}$} \\
\hline
Energy (GeV) & 45.60 & 182.50 \\
\hline
Horizontal tune $Q_x$ & 218.16 & 398.15\\
\hline
Vertical tune $Q_y$ & 222.20 & 398.22 \\
\hline
Horizontal emittance (nm) & 0.71 & 1.57 \\
\hline
Vertical emittance (pm) & 1.90 & 1.60 \\
\hline
$ \beta^*$ at IP x/y (mm) & 110 / 0.7 & 800 / 1.5 \\
\hline
Luminosity / IP ($\times 10^{34}  \mathrm{cm}^2 \mathrm{~s}$) & 141 & 1.38 
 \\
\hline
\end{tabular}
\\
\vspace{0.5cm} 

\label{tab:parameterstable}
\end{table}

Aiming for such ultra-low vertical emittance highlights the necessity for a comprehensive understanding of tolerance requirements regarding magnet field imperfections and misalignments, which will significantly impact the machine performance; large optics tuning simulation campaigns are needed to achieve this. 

Tuning simulations for the FCC-ee lattice at \( t\overline{t} \) energy, including a comprehensive correction strategy, were performed using the Methodical Accelerator Design software (MAD-X) \cite{charles23}.
In this paper, we focus on the FCC-ee baseline optics at Z energy, because although both energy modes are challenging to correct, the Z energy mode has shown to be more sensitive to errors compared to \( t\overline{t} \) \cite{thesis}
which makes this lattice well-suited for detailed and robust tuning studies, which can subsequently be applied to other modes. Additionally, the higher energy operation modes are scheduled for later stages in the FCC-ee physics program at CERN, giving higher priority to Z energy lattice analysis.

We used the Python interface for the Accelerator Toolbox (PyAT) \cite{pyAT} since it is a user friendly tool that leverages open-source scientific libraries in Python and provides needed support for users. By using PyAT we explored the potential of several correction algorithms for FCC-ee optics tuning. We developed our own correction procedure, tailored it to the highly-sensitive optics, and conducted comprehensive simulation studies.

\section{Correction algorithms}

The measurement and correction of beam orbit and optics have long been critical challenges, with numerous techniques developed and tested over decades to address them effectively \cite{phaseAdvance}. The aim of orbit and optics correction algorithms is to minimize impact of lattice errors by adjusting magnet strengths. The primary goal of these algorithms is to achieve optics parameters that closely resemble the design optics. This helps to recover the DA, momentum acceptance, and lifetime, while minimizing emittance and the beta function at the interaction point, thereby enhancing machine performance. 

 \subsection{SVD orbit correction}

Orbit correction involves generating Orbit Response Matrix (ORM)
for the closed orbit as a function of the orbit corrector kicks, the matrix is of dimension $m \times n$,  where $n$ is the number of used orbit correctors and $m$ is the number of BPMs. 

For the $i^{\textrm{th}}$ BPM and the $j^{\textrm{th}}$ corrector the ORM element is:

\begin{align}
M_{i,j} &= \frac{\sqrt{\beta_i(s) \beta_j(s_0)}}{2 \sin (\pi Q)} 
           \cos \left(\pi Q - \psi_i(s) + \psi_j(s_0)\right) \nonumber \\
        &\quad + \frac{\eta_i(s) \eta_j(s_0)}{\alpha_c L_o},
       \label{orm} 
\end{align}

\noindent where \( \beta \) is the beta functions at an element position, \( Q \) denotes the tune, \( \psi \) is the phase advance, \( \eta \) is the dispersion function, \( \alpha_c \) is the momentum compaction factor and \( L_o \) is the circumference of the accelerator.

The Singular value decomposition (SVD) orbit correction aims to invert the response matrix to find the proper orbit correctors kicks $\theta$ that best minimize the distortion on the closed orbit $\Delta x$, by following the relation $\ensuremath{\Delta x+M\Delta\theta=0}$, the SVD technique \cite{svd} is used to solve the least square problem. A proper selection for the orbit corrector kick used to generate the ORM has been discussed in Ref.\cite{ipac23}.

\subsection{LOCO for optics correction}

The Linear Optics from Closed Orbit (LOCO) technique \cite{loco} exploit the extensive information encoded in the ORM illustrated in Eq.~\ref{orm} by fitting measured
ORM $M$ to a lattice model $M_{model}$. The minimization process involves adjusting vector of parameters $p$ in iterations till the convergence of the correction of the optics parameters distortion such as the beta beating and relative dispersion, these parameters could be for example, the quadrupole gradients, quadrupole rolls (skew quadrupole gradients) in addition to BPMs and orbit correctors horizontal and vertical calibrations. Equation~\ref{deltah} illustrates the parameters update formula for the Gauss-Newton (GN) method adopted by the original LOCO \cite{thesis}:

\begin{equation}
\delta h_{\mathrm{GN}} = \left[J^{\top} W J\right]^{-1} J^{\top} W (M - M_{model}),
\label{deltah}
\end{equation}

\noindent where \(W\) is the diagonal weights matrix given by \(W = \frac{1}{\sigma^2}\), \(\sigma\) is the measured noise level on the BPMs (variance of input error) and $J = \left(\frac{\partial {M}}{\partial {p}}\right)$ is the Jacobian matrix, which gives an indication of how the function is sensitive to the change of its fitting parameters. The SVD is introduced here to solve the matrix inversion $ [\left[{J}^{\top} {W} {J}\right]]^{-1}$.

\subsection{Phase advance + $\eta_x$ correction}

Unlike optics correction methods based on the orbit response technique like LOCO, the Turn-By-Turn (TBT) and Multi-Turn techniques excite the beam and record beam position data over one or multiple turns to determine optics parameters like the betatron phase advance \cite{tbt}. Research has demonstrated that tuning the phase advance is equally effective as beta function correction \cite{phaseAdvance}, providing a powerful numerical method for linear optics correction by minimizing the difference between measured and model phase advances between adjacent BPMs. From the  lattice model, one can obtain the response matrix
$C$, that relates the relative phase advance $\delta \Delta {\psi}$ and relative horizontal dispersion $\delta \eta_x$ to the relative strengths $\delta {K}$ of all
quadrupole, as following:

\begin{equation}
\begin{pmatrix}
\alpha_1 \Delta {\psi}_x \\
\alpha_1 \Delta {\psi}_y \\
\alpha_2 {\eta}_x
\end{pmatrix}_{\text{meas}} = -C_{model} {\delta K},
\end{equation}

\noindent and then the SVD method is used 
to solve this system (similar to  orbit correction as described previously). The weights $\alpha_2=1-\alpha_1$ are introduced in order to determine the best compromise between correction of phase advance and dispersion.

\subsection{Coupling RDTs + $\eta_y$ correction}

Using Hamiltonian and normal form theory \cite{rdts}, \cite{rdts2} Resonance Driving Terms (RDTs) have been defined and it proved to be a powerful tool to describe betatron coupling close to the sum and difference resonance, and it is defined as following:

\begin{align}
    f_{\substack{1001 \\ 1010}}(s) &= -\frac{1}{4\left(1 - e^{2 \pi i\left(Q_x \mp Q_y\right)}\right)} \sum_l K^{sl}(s) \sqrt{\beta_x^l(s) \beta_y^l(s)} \nonumber \\
    &\quad \times e^{i\left(\Delta \psi_x^{sl}(s) + \Delta \psi_y^{sl}(s)\right)},
    \label{rdtsf}
\end{align}

\noindent where $ K^{sl}(s)$ is the $l^{\textrm{th}}$ integrated skew quadrupole strength, $\beta_{x,y}^l(s)$ are the beta functions at the location of the  $l^{\textrm{th}}$ skew quadrupole, $\Delta \psi_{x,y}^{sl}(s)$ are the phase advances between the observation point s and the $l^{\textrm{th}}$ skew quadrupole.

A relation detailed in Ref.~\cite{Zimmermann}, shows that, in order to minimize the vertical emittance, the driving terms $f_{\substack{1001 \\ 1010}}$, should be minimized. In order to correct coupling, skew quadrupoles must be installed, primarily at each sextupole magnet. Similar to betatron phase, the coupling matrix $N$ can be inferred from BPMs data. Minimization is applied to determine the appropriate skew quadrupole strength.

 The system to invert via SVD reads:

\begin{equation}
\begin{pmatrix}
\alpha_1 {f}_{1001} \\
\alpha_1 {f}_{1010} \\
\alpha_2 {\eta}_y
\end{pmatrix}_{\text{meas}} = -N_{model} {\delta K_s}.
\end{equation}

\section{Tuning simulation}

To conduct tuning simulation, we begin by installing one horizontal and one vertical orbit corrector at each of the lattice 1856 quadrupoles. Additionally, a BPM is placed at every quadrupole, including the final quadrupole doublets adjacent to the Interaction Points (IPs). To correct coupling and rematch the vertical dispersion, skew quadrupole correctors, are placed at each of the lattice 632 sextupoles. Our tuning simulations were based on some assumptions such that, emittances were calculated for non-colliding beams, meaning that beam-beam effects were not considered. Synchrotron Radiation (SR) effects were excluded from initial simulations, except for  emittance calculations, and its effect was addressed in a separate study. Additionally, solenoid errors were not included in the simulations.

\subsection{Optics sensitivity to magnet alignment errors}

We conducted a sensitivity analysis to investigate the impact of magnet alignments errors on key optics parameters. Sextupoles were set to 100$\%$ of their design value.
Horizontal and vertical random alignment errors distributed via a Gaussian
distribution with a cutoff at 2.5 times the standard deviation were systematically introduced first to the arc quadrupoles, sextupoles, and dipoles, and then to the IR magnets using standard deviations of (in micrometers): 
\begin{equation*}
    [0.0, 0.2,0.4, 0.6, 0.8, 1.0, 1.2 ].
\end{equation*}

These misalignment values are unrealistically small and are just used to enable insightful comparisons between the sensitivity to alignment errors of the arc and IR region separately. Fifty simulations were conducted, each employing different random seed, and the
assigned alignment errors followed the same standard deviation range.

\begin{figure}[h]
    \centering
    \begin{minipage}{0.45\textwidth}
        \centering
        \includegraphics[width=\textwidth]{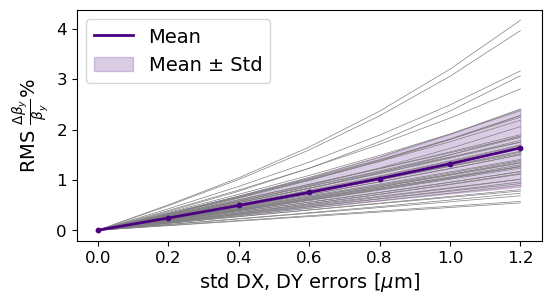}
        \label{fig:arc_optics}
    \end{minipage}\hfill
    \begin{minipage}{0.45\textwidth}
        \centering
        \includegraphics[width=\textwidth]{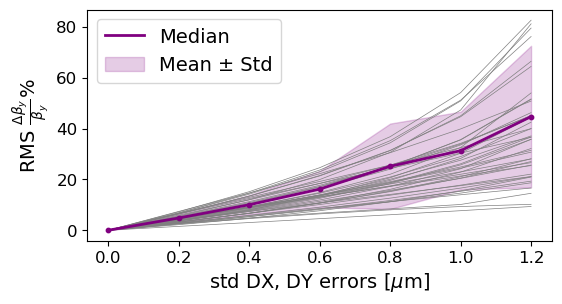}
        \label{fig:second_figure}
    \end{minipage}
    \caption{Vertical percentage beta beating vs. error standard deviation for the arc region (upper) and IR (lower).}
    \label{fig:errorVsoptics}
\end{figure}

Figure~\ref{fig:errorVsoptics} illustrates an example of the resulting plot from the conducted analysis. The figure shows that applying 1 \(\mu\text{m}\) errors led to a vertical percentage beta beating with a mean value (averaged over 50 seeds) of about 1.4\% when errors are applied to the arc magnets, whereas the mean value increased to about 30\% when errors are applied to the IR magnets. This greater sensitivity of the IR magnets compared to the arc magnets was consistent across all optics parameters, including closed orbit dissociation and relative dispersion, and was most pronounced for the vertical emittance \cite{thesis}. The increased sensitivity of the IR magnets to alignment errors is due to the stronger magnets used to focus the beam and correct chromaticity at the IP. This greater sensitivity introduces tuning simulation challenges in that region, which will likely be addressed by a dedicated alignment system, with feasible tolerances provided by the FCC-ee magnet alignment group to guide future tuning simulations.

\subsection{Correction procedure}

The baseline lattice was used at full beam energy of 45.6 GeV. The correction procedure is as follows:

\begin{enumerate}
    \item \textbf{Initial setup:} At the start of the simulation, radiation effects are turned off and nonlinear effects from sextupoles are initially neglected by turning off the sextupoles. This allows us to start with a linear lattice that can tolerate larger imperfections. This approach is standard in the commissioning of fourth-generation light sources \cite{Sajaev}.
    
    \item \textbf{Alignment errors and initial corrections:} Alignment errors are applied to the arc magnets. This is followed by beam threading and initial orbit correction and tune fitting, which utilizes all the focusing and defocusing quadrupoles in the arc region.

    \item \textbf{Sextupole ramping:} At this stage, sextupole fields need to be switched on again to manage the chromatic effects of the lattice and dampen head-tail instabilities \cite{chrom}. We ramp up sextupole strength in 10\% increments, with further orbit and tune corrections performed at each step to avoid resonances and ensure overall beam stability.

    \item \textbf{Chromaticity correction:} Once sextupole strengths reach 100\% of the design value, the arc focusing and defocusing sextupole strengths are varied to perform chromaticity correction.

    \item \textbf{Optics correction:} At this stage the beam lifetime is long enough to start detailed optics correction steps, including beta beating, coupling, and horizontal and vertical dispersion corrections.
\end{enumerate}

\begin{figure}[htbp!]
   \centering
   \includegraphics[width=0.5\textwidth]{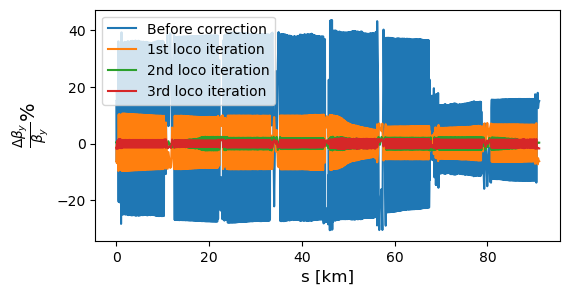}
   \caption{Improvement of vertical beta beating across three LOCO iterations.}
   \label{fig:loco}
\end{figure}

We implemented Python-based numerical code for LOCO \cite{ipac23}. This implementation includes numerical calculations of ORMs and Jacobians, and the design of LOCO iteration strategies with choosing the proper parameters values. The validity of the implemented code is demonstrated in Fig.~\ref{fig:loco} where random relative field errors, distributed via a Gaussian distribution with a standard deviation of \(\Delta k/k = 2 \times 10^{-4}\), were applied to all quadrupoles and sextupoles. The figure shows the reduction in vertical beta beating over each LOCO iteration until convergence.

For FCC-ee LOCO correction iterations are interleaved with orbit correction to prevent emittance growth.
The described correction procedure was utilized, with horizontal and vertical random alignment errors (with a standard deviation of 100 $\mu$m) applied to arc magnets. Initially, LOCO utilised all normal quadrupoles to correct the beta beating. However, the tuning simulation resulted in a mean vertical emittance value of 5.99 pm, which exceeded the design value for the lattice (0.7 pm, excluding the beam-beam effect). Therefore, we followed LOCO with coupling RDTs correction, which helped in achieving a vertical emittance of 0.75 pm. The achieved DA after correction is shown in Fig.~\ref{fig:arc_optics}.

\begin{figure}[htbp!]
    \centering
    \includegraphics[width=0.45\textwidth]{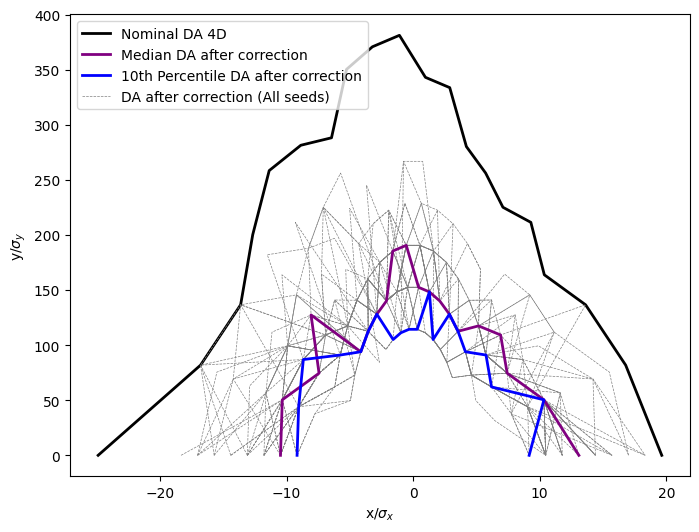}
    \caption{DA after correction using LOCO with all normal quadrupoles followed by RDTs + $\eta_y$ correction.}
    \label{fig:arc_optics}
\end{figure}

\begin{figure}[htbp!]
    \centering
    \includegraphics[width=0.45\textwidth]{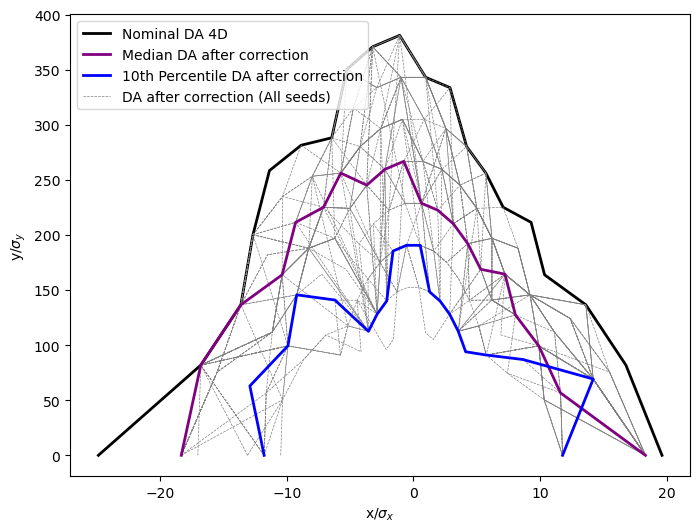}
    \caption{DA after correction using LOCO with normal trim quadrupoles at sextupoles followed by RDTs + $\eta_y$ correction.}
    \label{fig:arc_optics2}
\end{figure}

We examined different corrector locations by using normal trim quadrupole components installed at every sextupole magnet to correct the beta beating. The new set of correctors showed an improvement in the achieved DA, as shown in Fig.~\ref{fig:arc_optics2}.

\subsection{Phase advance for beta correction}

We participated in a benchmarking study of commissioning simulations with errors and corrections between AT and MAD-X to understand and correct discrepancies between the two computer codes \cite{Benchmark}. The study demonstrated good agreement between the two codes when the comparison were performed up to orbit correction step and the agreement between the codes suggests that different tuning simulation results are due to different optic correction algorithms. As outcome of the benchmarking study we started to explore different optics correction algorithms by integrating phase advance and horizontal dispersion correction into our scheme instead of LOCO to correct the beta beating and horizontal dispersion.

\begin{figure}[htbp!]
    \centering
    \includegraphics[width=0.45\textwidth]{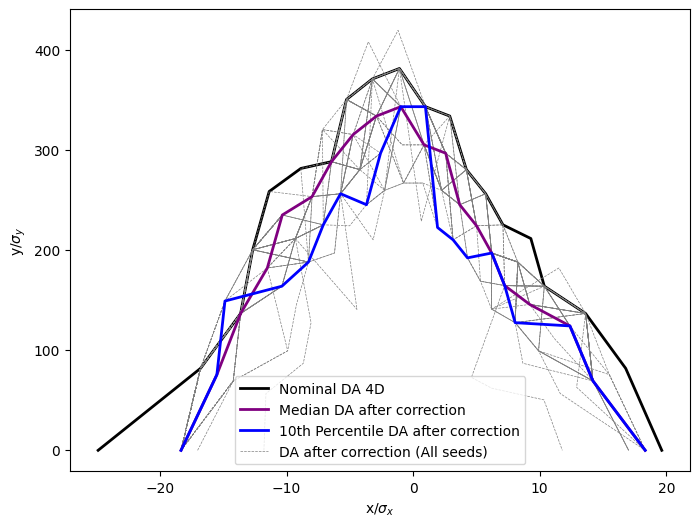}
    \caption{DA after correction of random horizontal and vertical alignment errors with a standard deviation of 100 $\mu$m on the arc components of the baseline lattice, using phase advance + $\eta_x$ and RDTs + $\eta_y$ corrections, for 50 seeds.}
    \label{fig:DA_phase}
\end{figure}

Figure~\ref{fig:DA_phase} illustrates the achieved DA using phase advance + $\eta_x$ and RDTs + $\eta_y$ optics correction instead of LOCO. The achieved median DA of the seeds when applying phase advance + $\eta_x$ and RDTs + $\eta_y$ covers an area range from -20 to 20 horizontally and up to about 320 vertically in multiples of the horizontal and vertical beam sizes, the larger median DA compared to LOCO DA in Fig.~\ref{fig:arc_optics2} highlights a better performance of the phase advance + $\eta_x$ and RDTs + $\eta_y$ corrections for the lattice and conditions we examined.

\begin{figure}[htbp!]
    \centering
    \subfigure[Vertical emittance frequency distributions]{
        \includegraphics[width=0.45\textwidth]{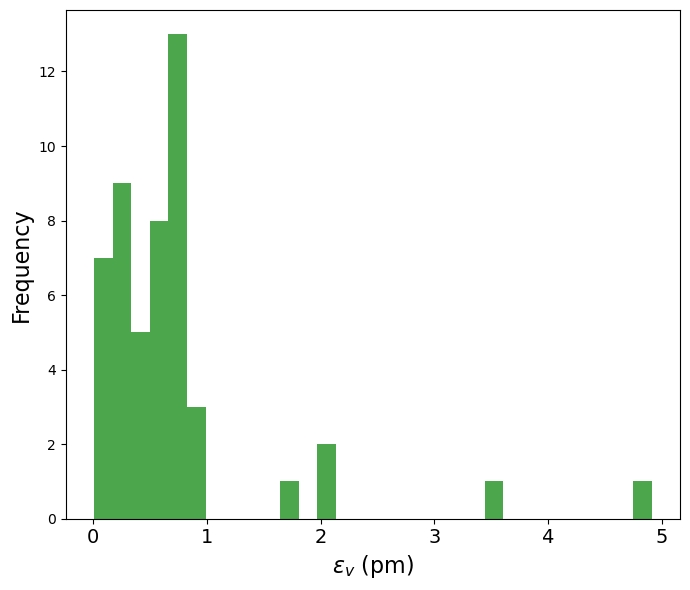}
        \label{fig:emit_phase}
    }
    \subfigure[Horizontal emittance frequency distributions]{
        \includegraphics[width=0.45\textwidth]{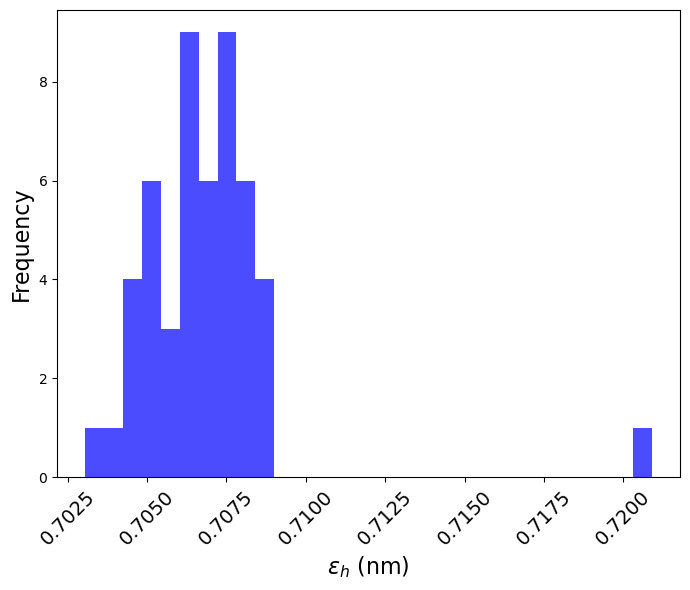}
        \label{fig:emit_phase2}
    }
    \caption{Frequency distributions of vertical and horizontal emittance after applying the correction procedure for 50 seeds. Horizontal and vertical misalignments with a standard deviation of 100 $\mu$m were applied to all arc magnets.}
    \label{fig:emit_distributions}
\end{figure}

A histogram of the vertical emittance after correction, shown in Fig.~\ref{fig:emit_phase}, indicates that the corrected seeds primarily achieved vertical emittance values below 1 pm, with only a few seeds exceeding this threshold. Similarly, the horizontal emittance distribution, shown in Fig.~\ref{fig:emit_phase2}, demonstrates that none of the seeds surpassed the design value.

\subsection{Tuning simulations with SR}

The effectiveness of the developed tuning tools in the presence of Synchrotron Radiation (SR) was examined. Including the SR in PyAT was done by activating longitudinal motion which by default, enables radiation effects in the elements, activates RF cavities, and collective effects. Subsequently, additional correction steps were introduced into the correction procedure, including adjusting the proper RF cavity phase and frequency, and performing tapering to compensate
the large energy loss per turn caused by SR and the large saw-tooth distortion in the horizontal orbit.

The tuning simulation including optics correction with phase advance + $\eta_x$ and coupling RDTs + $\eta_y$ was performed by assigning the 
values illustrated in Tab.~\ref{tab:displacementTiltTable}. The applied errors were randomly distributed via a Gaussian distribution truncated at 2.5 sigma.

\begin{table}[htbp!]
\caption{Introduced misalignment errors prior to applying corrections.}
\centering
\small
\setlength{\tabcolsep}{4pt} 
\renewcommand{\arraystretch}{1.2} 
\begin{tabular}{|c|c|c|}
\hline
\textbf{Element} & \textbf{$\sigma_{x/y}$ ($\mu$m)} &  \textbf{$\sigma_{\theta}$ ($\mu$rad)}  \\
\hline
Arc quads \& sext. & 100  & 100 \\
\hline
Dipoles & 1000  & 1000 \\
\hline
BPMs & Same as quads &  - \\
\hline
\end{tabular}
\\
\vspace{0.5cm} 

\label{tab:displacementTiltTable}
\end{table}

We conducted a study that have shown a preference for aligning the BPMs to quadrupoles rather than to sextupoles \cite{tuningMeeting}, specially when considering the girders misalignments. The offset values for these BPMs were set to match the assigned offset errors of the corresponding quadrupoles, assuming an ideal scenario where the random misalignments of the BPMs precisely match those of the attached quadrupoles. Consequently, Beam Based Alignment (BBA) procedure is not required.

\begin{figure}[htbp!]
    \centering
    \begin{minipage}{0.45\textwidth}
        \centering
        \includegraphics[width=\textwidth]{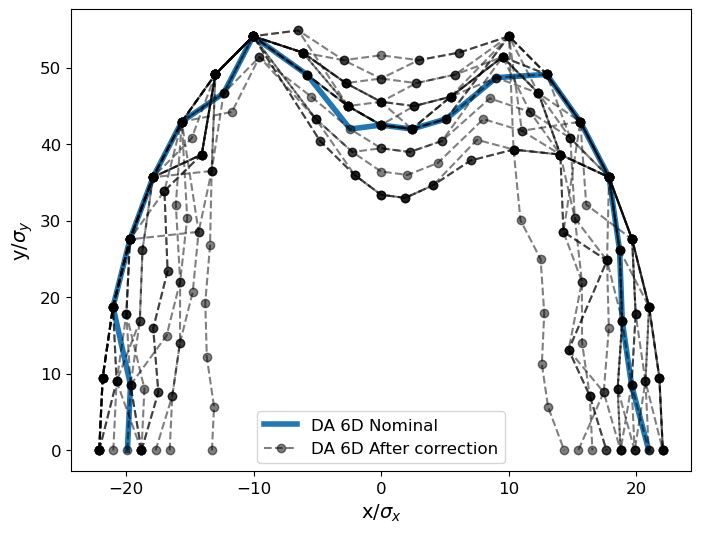}
        \label{fig:bpm_to_q}
    \end{minipage}\hfill
    \begin{minipage}{0.45\textwidth}
        \centering
        \includegraphics[width=\textwidth]{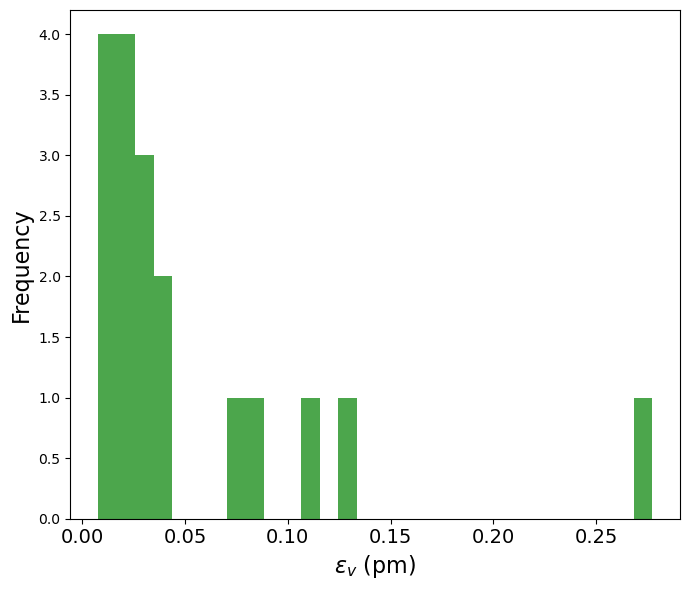}
        \label{fig:second_figure}
    \end{minipage}
    \caption{DA (upper) and vertical emittance  frequency (lower) after correction including SR and BPMs aligned to quadrupoles (with one removed large outlier)}
    \label{fig:bpmtoq}
\end{figure}

Figure~\ref{fig:bpmtoq} shows the DA and vertical emittance distribution after correction of 20 seeds. One seed failed to complete the simulation, and another seed resulted in a large vertical emittance of 300 pm and so was considered to have failed. The achieved mean vertical emittance was 0.054 pm for 18 seeds.

\subsection{Girders misalignments}

\begin{figure}[htbp!]
    \centering
    \includegraphics[width=0.45\textwidth]{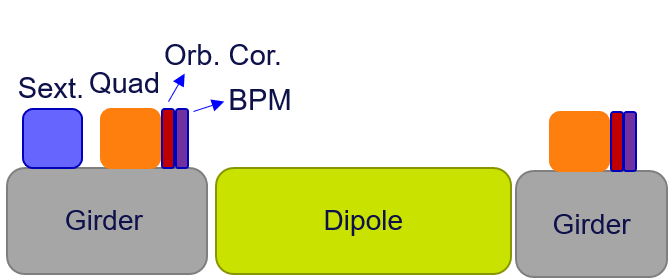}
    \caption{Girders configuration for section of the arc region}
    \label{fig:girder}
\end{figure}

Figure~\ref{fig:girder}
 illustrates the girders configuration under consideration. These girders support both sextupoles and quadrupoles positioned between two dipoles in the arc region, or quadrupoles that are not adjacent to sextupoles. The proposed alignment between girders has a tolerance of $\sigma$ = 150 $\mu$m, while the alignment between magnets is set to $\sigma$ = 50 $\mu$m. The assigned alignment error values, including those for the girders, are detailed in Tab.~\ref{tab:displacementTiltTable3}.

\begin{table}[htbp!]
\caption{Introduced misalignment errors including girders.}
\centering
\small
\setlength{\tabcolsep}{4pt} 
\renewcommand{\arraystretch}{1.2} 
\begin{tabular}{|c|c|c|}
\hline
\textbf{Element} & \textbf{$\sigma_{x/y}$ ($\mu$m)} &  \textbf{$\sigma_{\theta}$ ($\mu$rad)}  \\
\hline
Arc quads \& sext. & 50  & 50 \\
\hline
Dipoles & 1000  & 1000 \\
\hline
Girders & 150 &  - \\
\hline
BPMs & Same as quads &  - \\
\hline
\end{tabular}
\vspace{0.5cm} 
\label{tab:displacementTiltTable3}
\end{table}

\begin{figure}[htbp!]
    \centering
    \begin{minipage}{0.45\textwidth}
        \centering
        \includegraphics[width=\textwidth]{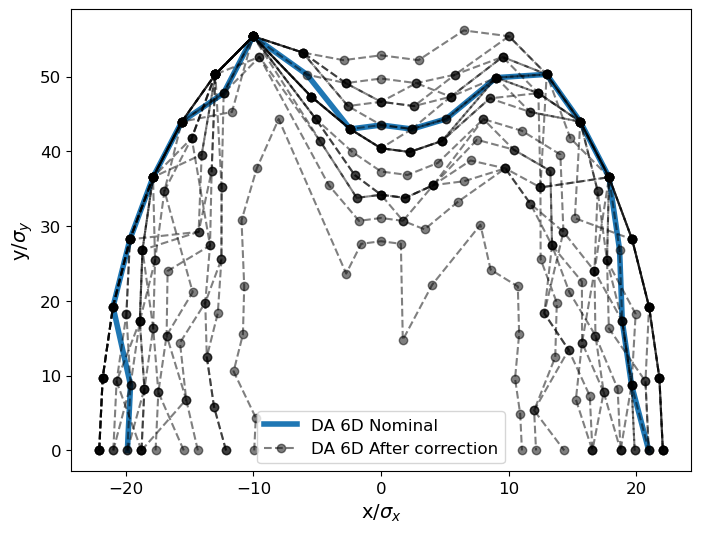}
        \label{fig:bpm_to_q}
    \end{minipage}\hfill
    \begin{minipage}{0.45\textwidth}
        \centering
        \includegraphics[width=\textwidth]{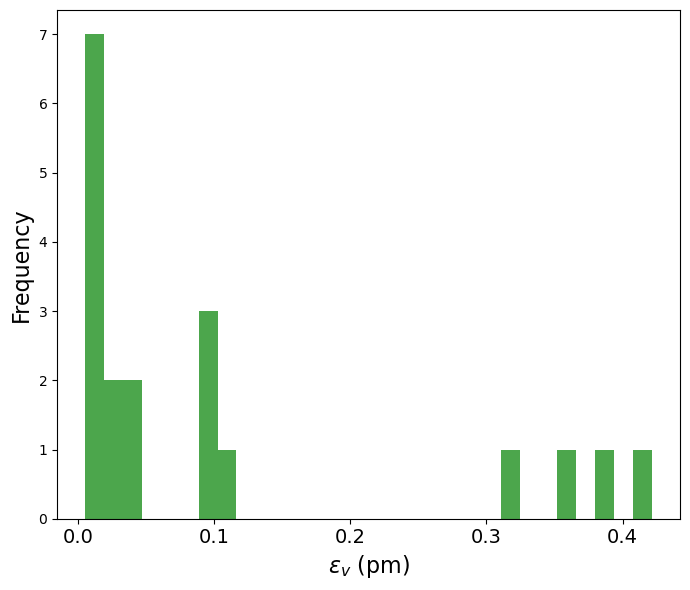}
        \label{fig:second_figure}
    \end{minipage}
    \caption{DA (upper) and vertical emittance  frequency (lower) after correction including girders misalignment}
    \label{fig:tuninggirders}
\end{figure}

The achieved optics parameter values after following the correction procedure are shown in Tab.~\ref{tab:optics} where the 20 seeds managed to achieve median vertical emittance value of 0.04 pm. The vertical emittance histogram and the achieved DA after correction are shown in Fig.~\ref{fig:tuninggirders}, the figures show that including girder misalignments has added to the challenges of the simulation in terms of the achieved DA for some seeds, however all the seeds managed to achieved the design vertical emittance value.

\begin{table}[htbp!]
    \centering
    \vspace{0.5cm} 

    \caption{Median values of several optics parameters after correction.}
    \label{tab:optics}
    \begin{tabular}{|c|c|}
        \hline
        \textbf{Parameter} & \textbf{Value} \\
        \hline
        \text{rms hor. orbit ($\mu$m)} & 38.94 \\
        \hline
        \text{rms ver. orbit ($\mu$m)} & 37.16 \\
        \hline
        \text{rms $\Delta\beta_x/\beta_x$ (\%)} & 1.32 \\
        \hline
        \text{rms $\Delta\beta_y/\beta_y$ (\%)} & 12.73 \\
        \hline
        \text{rms $\Delta\eta_x$ (mm)} & 14.78 \\
        \hline
        \text{rms $\Delta\eta_y$ (mm)} & 0.89 \\
        \hline
        \text{$\varepsilon_h$ (nm)} & 0.71 \\
        \hline
        \text{$\varepsilon_v$ (pm)} & 0.04 \\
        \hline
        \text{rms hor. $\Delta\psi$} & $1.45 \times 10^{-3}$ \\
        \hline
        \text{rms ver. $\Delta\psi$} & $1.04 \times 10^{-2}$ \\
        \hline
        \text{rms Re F1001} & $2.89 \times 10^{-4}$ \\
        \hline
        \text{rms Im F1001} & $2.65 \times 10^{-4}$ \\
        \hline
        \text{rms Re F1010} & $2.96 \times 10^{-4}$ \\
        \hline
        \text{rms Im F1010} & $2.96 \times 10^{-4}$ \\
        \hline
    \end{tabular}
\end{table}

\subsection{Tuning simulation for ballistic optics}

\begin{figure}[htbp!]
    \centering
    \begin{minipage}{0.45\textwidth}
        \centering
        \includegraphics[width=\textwidth]{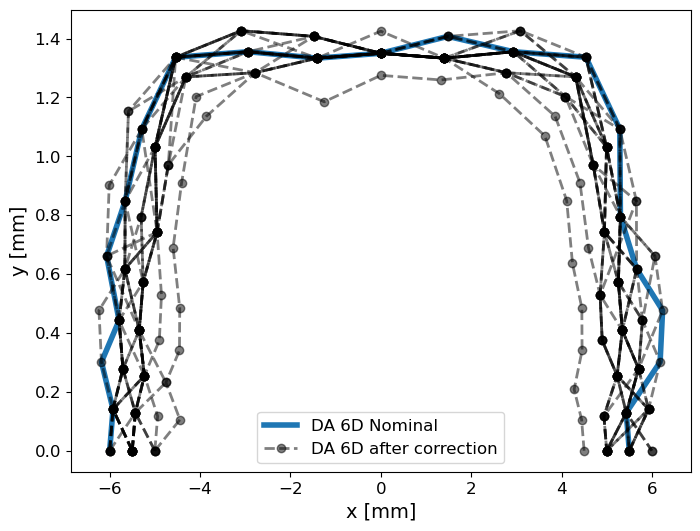}
        \label{fig:DA_ball_final}
    \end{minipage}\hfill
    \begin{minipage}{0.45\textwidth}
        \centering
        \includegraphics[width=\textwidth]{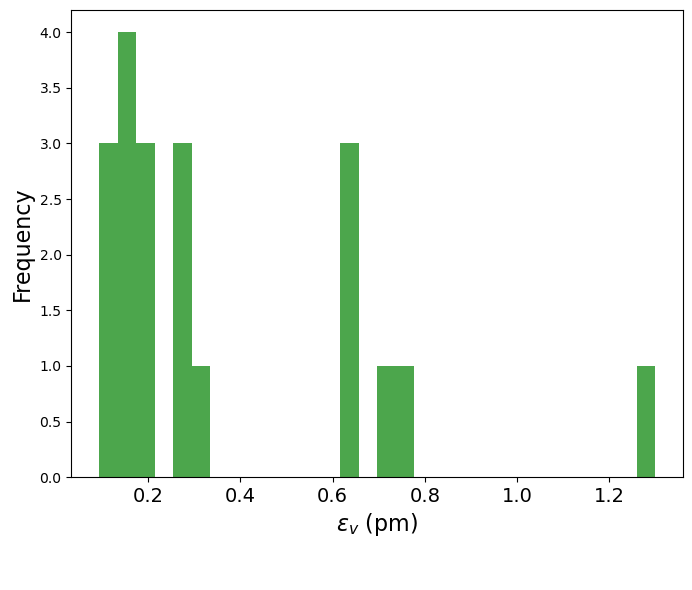}
        \label{fig:bals_final}
    \end{minipage}
    \caption{DA (upper) and vertical emittance  frequency (lower) after correction of Ballistic optics}
    \label{fig:DAballistics}
\end{figure}

Tuning simulations have been performed for the ballistic commissioning optics \cite{ballistic}, that involves turning off certain IR quadrupoles and all IR sextupoles, thereby facilitating optics tuning. In these simulations, we started to encounter BPM-to-quadrupole alignment errors, these values are initially set to $\sigma = 100\ \mu$m. Subsequent BBA simulations were performed \cite{BBA}, and showed that these values were reduced to 10 $\mu$m, as shown in Tab.~\ref{tab:balisticTable}. Median values for optics parameters after correction are illustrated in Tab.~\ref{tab:balisticOptics}. Figure~\ref{fig:DAballistics} shows the DA and vertical emittance distribution after correction.

\begin{table}[htbp!]
\caption{Introduced misalignment errors to Ballistic optics.}
\centering
\small
\setlength{\tabcolsep}{4pt} 
\renewcommand{\arraystretch}{1.2} 
\begin{tabular}{|c|c|c|}
\hline
\textbf{Element} & \textbf{$\sigma_{x/y}$ ($\mu$m)} &  \textbf{$\sigma_{\theta}$ ($\mu$rad)}  \\
\hline
Arc quads \& sext. & 50  & 50 \\
\hline
Dipoles & 1000  & 1000 \\
\hline
Girders & 150 &  150 \\
\hline
BPMs-to-quads & 10 &  - \\
\hline
\end{tabular}
\vspace{0.5cm} 
\label{tab:balisticTable}
\end{table}

\begin{table}[htbp!]
    \centering
    \vspace{0.5cm} 

    \caption{Median values of optics parameters after correction for Ballistic optics.}
    \label{tab:balisticOptics}
    \begin{tabular}{|c|c|}
        \hline
        \textbf{Parameter} & \textbf{Value} \\
        \hline
        \text{rms hor. orbit ($\mu$m)} & 130.36 \\
        \hline
        \text{rms ver. orbit ($\mu$m)} & 144.75 \\
        \hline
        \text{rms $\Delta\beta_x/\beta_x$ (\%)} & 1.02 \\
        \hline
        \text{rms $\Delta\beta_y/\beta_y$ (\%)} & 0.63 \\
        \hline
        \text{rms $\Delta\eta_x$ (mm)} & 0.66 \\
        \hline
        \text{rms $\Delta\eta_y$ (mm)} & 1.68 \\
        \hline
        \text{$\varepsilon_h$ (nm)} & 0.85 \\
        \hline
        \text{$\varepsilon_v$ (pm)} & 0.23 \\
        \hline
        \text{rms hor. $\Delta\psi$} & $1.15 \times 10^{-3}$ \\
        \hline
        \text{rms ver. $\Delta\psi$} & $1.12 \times 10^{-3}$ \\
        \hline
    \end{tabular}
\end{table}

 \section{CONCLUSION AND OUTLOOK}

In this paper, we addressed the alignment error tolerances of the arc magnets in the proposed FCC-ee baseline optics at the Z energy. We evaluated the effectiveness of the optics and orbit correction algorithms implemented in PyAT through a tuning scheme designed to optimize machine performance. A comparison between LOCO and phase advance optics correction  demonstrated superior performance for the latter, resulting in an enhanced DA. With the current procedure, for the nominal lattice we achieved a median vertical emittance of 0.04 pm and a horizontal emittance of 0.71 nm when applying 150 $\mu$m shift errors to the arc magnet girders, 50 $\mu$m alignment errors on arc quadrupoles and sextupoles and 1000 $\mu$m on all dipoles. For the ballistic optics, the achieved median vertical emittance is 0.23 pm. The simulations for the ballistic optics account for a BPM-to-quadrupole misalignment of 10 $\mu$m after BBA and include a 150 $\mu$m tilt added to the girders. Looking forward, we plan to conduct tuning simulations under more realistic conditions, incorporating magnet strength errors, IR magnet errors, and multipolar errors. Investigating these factors and exploring advanced correction algorithms and computational techniques, aim to further optimize the performance of the FCC-ee and future fourth-generation light sources.

\section{ACKNOWLEDGEMENTS}

This work was supported by the European Union’s Horizon 2020 research and innovation programme under grant agreement No 951754, we also wish to thank Katsunobu Oide, Frank Zimmermann, Rogelio Tomas, Simone Liuzzo and Simon White for valuable insights and contributions.

\end{document}